# Hedge Fund Index Rules and Construction

David Xiao

**ABSTRACT**

A Hedge Fund Index is very useful for tracking the performance of hedge fund investments, especially the timing of fund redemption. This paper presents a methodology for constructing a hedge fund index that is more like a quantitative fund of fund, rather than a weighted sum of a number of early replicable market indices, which are re-balanced periodically. The constructed index allows hedge funds to directly hedge their exposures to index-linked products. That is important given that hedge funds are an asset class with reduced transparency, and the returns are traditionally difficult to replicate using liquid instruments.

Hudge funds have been growing rapidly over the last decades due to their important roles in financial markets. Hedge funds can be regarded as unregulated mutual funds that are not publicly traded in organized exchanges. The return and risk of a hedge fund are quite different from traditional investments, which makes hedge funds highly volatile.

Hedge fund indices are designed to capture the performance and behavior of most liquid hedge funds. It appears reasonable to describe hedge fund performance based on a portfolio of hedge funds that provides a comprehensive spectrum of possible market exposures. Indices are designed to capture the general direction and size of market movements and are generally accepted as good proxies for overall market behavior of a given sector.

There is rich literature on hedge fund index. Atilgan et al (2013) investigate the performance of various hedge fund indices. They find that the risk-adjusted performance of most hedge fund indices deteriorates over time and conclude that non-investable indices are superior performers with respect to their investable counterparts. Kapil and Gupta (2019), Dimmock et al (2016) account for various hedge fund indices that have performed in the past and find that most hedge fund indices witness a drop in returns over time, and most hedge funds do not provide additional diversification benefits comparing to traditional asset class.

Hwang et al (2017) examine the cross-sectional relation between hedge fund returns and systemic risk and find evidence for a positive and statistically significant relation between systemic risk and hedge fund returns. Ardia and Boudt (2018) define an outperformance measure of hedge funds and show that the p-values of pairwise tests of equal performance can be used to obtain estimates of the out- and under performance ratio that are robust to false discoveries.

Beschwitz et al (2022) exploit transaction and position data for a sample of long-short equity hedge funds to study the trading activity of fundamental investors and find that hedge funds close their positions too early, thereby forgoing about one-third of the trades' potential profitability. Stafylas and Anderson (2018) examine hedge fund index construction methodologies, by describing and analyzing case studies and evaluating them using numerical examples.

Barth et al (2021) estimate that the worldwide net assets under hedge fund management is larger than the most generous estimate and show that the total returns earned by funds that report to the public databases are significantly lower than the returns of funds that report only on regulatory filings. Joenväärä et al (2019) examine the fundamental questions regarding hedge fund performance and find a significant association between fund-characteristics related to share restrictions as well as compensation structure and risk-adjusted returns.

An index may significantly increase the number of funds in order to meet notional inflow requirements. In such a situation there may be some difficulty finding funds that meet the current eligibility requirements, or the new funds that meet the requirements may not be of a particularly high 'quality'.

While the index does not aim to choose funds that are 'good performers', it does aim to track the industry. If a significant number of funds enter the index that do not faithfully represent their strategy identifications, then this could cause the index not to track the market as well as it should.

This paper proposes a new model for constructing hedge fund index. The constructed index runs more like a fund of fund. As a result, the fund can reflect the performance of hedge fund industry very well.

The model can select funds across different strategies. If there is an insufficient number of eligible funds in a given strategy to allocate the computed reallocation adjustments, the model can distribute the amount from other strategies to the current strategy.

We get hedge fund data from a proprietary data set, which is constructed from a number of vendor-supplied hedge fund databases. The funds appearing in this data set make up 'The Universe' of hedge funds as far as the index is concerned.

Since the index aims at being broadly representative of the hedge fund industry, it is important that The Universe represents a large enough fraction of the actual hedge fund universe.

Our index construction also ensures that non-Index-related hedge fund exposures elsewhere in the market can affect the details of how the index runs–for example, through fund weight caps. The fund of fund clearly aims at selecting funds that will perform well and will take up capacity in these funds. As the index attempts to increase its exposure, it will be capped in funds that perform well.

The rest of this paper is organized as follows: The model is described in Section 1; Section 2 discusses index weighting. Index adjustment is presented in Section 3; the conclusions are given in Section 4.

## 1. Hedge Fund Index Model

Although hedge fund indices are well known for reflecting the performance of the hedge fund industry, the actual operation of the index for tracking outstanding redemptions, expected client inflow and outflow, managing exposure limits, etc. is fairly complicated.

Mathematically, the basic relationship between index level and return is

$$\lambda_{m+1} = \lambda_m(1 + \kappa_m) \tag{1}$$

Where $\lambda_m$ is the index level at the beginning of a month m or the end of month m – 1, and $\kappa_m$ is the *fee-adjusted* index return for month m:

$$\kappa_m = \sum_i \omega^{b^i}_m \rho^i_m - \phi. \tag{2}$$

Here φ = 95bp/12 = 7.917bp is the monthly fee charged against the index, $\rho_m^i$ is the return of fund i in month m, and $\omega_m^{bi}$ is the weight of fund i at the beginning of month m. However, if we are in the middle of a month, the accrued fee is used instead:

$$\phi = \frac{\text{days into month}}{\text{total days in month}} \times \frac{1}{12} \times 95\text{bp}. \qquad (3)$$

So far it can be seen that the index return is the weighted sum of fund returns, which is then adjusted by the fee.

Although transparent, the above calculation is not as simple as it appears. In order to calculate the fund return $\rho_m^i$, we need to know the net asset value (NAV) of the fund in question at the beginning and end of the month. Typically, we do not have finalized fund NAVs until some weeks have passed, although we may have preliminary estimates.

This results in the index return being recalculated at various times with different estimates of the fund returns, until the finalized value of the index is calculated: 45 calendar days after the end of the month. Even then there may be some funds that have not reported finalized NAVs, and the index administrator may have estimated the return.

In addition, redemption fees may be charged by the fund for index-related adjustment activities, and these fees are 'charged' to the fund return. That is, if there is a redemption from a fund that would result in a fee, the fraction of that redemption attributable to index adjustment activities is used to adjust the fund return in the index.

Also, if the fund weight for the month consists partly of a 'residual weight' or 'passive weight', the residual weight fraction gets no performance. The residual weight can be thought of as an outstanding redemption from a fund that gets no market performance since it is no longer 'in' the fund, but it has either not yet been received or has not yet been re-invested or redeemed to a client.

In summary, the return would be calculated something like:

$$\rho = \frac{\omega - \zeta}{\omega} r, \qquad (4)$$

where $\omega$ is the total fund weight for the month and fund in question and $\zeta$ the residual weight for the fund. Here $r$ is the fee-adjusted fund return, which is currently determined on an ad hoc basis by the Index administrator.

Once the fund weight at the beginning of a month $\omega_m^b$ and its return are known, the weight at the end of the month is also calculated in the standard manner:

$$\omega_m^{ei} = \frac{1 + \rho_m^i}{1 + \kappa_m + \phi} \omega_m^{bi}. \qquad (5)$$

This is easy to understand: The weights must be normalized at all times: $\sum_i \omega_m^{ei} = 1 = \sum_i \omega_m^{bi}$ Pi and (5) is therefore consistent with the definition of the index return κ (2). The numerator: $(1 + \rho)\omega$ represents the performance of the fund, whereas the denominator $(1 + \kappa_m + \phi)$ is the normalization factor to scale the weights back to a fractional representation of the index.

To be more explicit, define (ignoring the distinction between $\omega_m^e$ and $\omega_{m+1}^b$ for the moment):

$$\mu_m^i = \lambda_m \omega_m^i, \quad \mu_{m+1}^i = \lambda_{m+1} \omega_{m+1}^i, \tag{6}$$

which represents the 'amount' of the index that a fund represents, rather than the fraction (or if multiplied by the total notional, it can be thought of as the dollar investment in the fund). In particular: $\lambda_m = \sum_i \mu_m^i$

The returns of these fund amounts that ignore the index fee are:

$$\tilde{\mu}_{m+1}^i = (1 + \rho_m^i)\mu_m^i, \tag{7}$$

which makes up the index level at the end of the month prior to fee: $\tilde{\lambda}_m = \sum_i \tilde{\mu}_{m+1}^i$

By (1) we must also have the new index level that has no fee as:

$$\tilde{\lambda}_{m+1} = (1 + \kappa_m + \phi)\lambda_m = \frac{(1 + \kappa_m + \phi)}{(1 + \kappa_m)}\lambda_{m+1}, \tag{8}$$

and multiplying each side by $\omega^i_{m+1}$ gives a relationship between the pre/post fee amounts:

$$\tilde{\mu}^i_{m+1} = \frac{(1 + \kappa_m + \phi)}{(1 + \kappa_m)}\mu^i_{m+1}. \tag{9}$$

Unsurprisingly, this is exactly the same as the relationship between the pre/post fee index levels above.

Combining this with (7) gives:

$$\frac{(1 + \kappa_m + \phi)}{(1 + \kappa_m)}\mu^i_{m+1} = (1 + \rho^i_m)\mu^i_m, \tag{10}$$

and then using (6) on both sides, then using (1) to cancel the index levels, gives:

$$(1 + \kappa_m + \phi)\omega^i_{m+1} = (1 + \rho^i_m)\omega^i_m, \tag{11}$$

which gives (5).

## 2. Hedge Fund Index Weights

The reason for the distinction between, for example, $\omega^e_m$ and $\omega^b_{m+1}$ is that the 'adjustments' to the index, i.e., re-balancing, removing ineligible funds, investing inflow, etc., are considered to be made between the end of the month and the beginning of the next month. The end result of making these adjustments will be a relationship between the fund weights at the end of the month and the fund weights at the beginning of the next, resulting from application of the rules. This is a multi-step process.

The 'Preliminary fund weight' is determined as:

$$\phi^{bi}_{m+1} = (\omega^{ei}_m - \zeta^{ei}_m)\frac{\eta^e_m}{\eta^b_{m+1}} + \sum_{k=1}^{m} \frac{\alpha^i_{m+1,k}\eta^b_{m+1,k}}{\eta^b_{m+1}} + \delta^{bi}_{m+1}. \quad (12)$$

To understand this, first note that $\eta_m$ is the net notional amount of index-related products outstanding in months m with beginning or end specified as before. This means that, for example, $\omega^{bi}_m \eta^b_m$ represent the $-amount of investment in fund i related to the index.

So then multiplying (17) by $\eta^b_{m+1}$ makes it slightly easier to understand:

$$\phi^{bi}_{m+1}\eta^b_{m+1} = (\omega^{ei}_m - \zeta^{ei}_m)\eta^e_m + \sum_{k=1}^{m} \alpha^i_{m+1,k}\eta^b_{m+1,k} + \delta^{bi}_{m+1}\eta^b_{m+1}. \quad (13)$$

The left side is an estimate of the $-amount that will be invested in fund i after adjustments. The first term on the right side represents the 'active' (or actual) investment in fund i, since ω is the total weight and ζ represents the residual weight at the end of the month. The last term on the right represents the calculated outstanding residual amounts for month m + 1, which have been calculated as a fraction of the notional at the beginning of month m + 1. So first and last terms are: the active weight remaining in the fund from the previous month plus the passive weights for the next month.

The middle term represents all adjustments that have been made that will become effective at the beginning of month m + 1. The reason that there is a sum is that these adjustments may have been calculated at different times in the past, represented by the summation index k. The adjustments are represented by an adjustment weight α, but the weight refers to an estimate of the notional amount in month m + 1 which was estimated in month $k$: $\eta^b_{m+1,k}$. The combination αη is therefore the $-amount of the adjustment make in month k that becomes effective in month m + 1.

To summarize, (17) says:

$$\text{estimated weight} = \text{previous active weight} + \text{new adjustments}$$
$$+ \text{passive weights for next month} \tag{14}$$

The reason that φ in (17) is called a preliminary weight, is that the new weights are not always properly normalized. This is counterintuitive, as one wouldn't call something a 'weight' if it wasn't normalized. Specifically, the application merely calculates the total weight and re-scales to get the fund weights:

$$\omega^{b^i}_{m+1} = \gamma_m \phi^{b^i}_{m+1}, \quad \gamma_m = 1/\sum_j \phi^{b^j}_{m+1}, \qquad (15)$$

which guarantees that $\sum_i \omega^{b^i}_{m+1} = 1.$

It is not difficult to derive a condition on the adjustments such that the normalization is automatic by summing (17) over all funds:

$$\eta^b_{m+1} - \eta^e_m = \sum_i \left( \delta^{b^i}_{m+1} \eta^b_{m+1} - \zeta^{ei}_m \eta^e_m \right) + \sum_i \sum_{k=1}^{m} \alpha^i_{m+1,k} \eta^b_{m+1,k} \qquad (16)$$

Note that the left side of this represents the notional in-/out-flow in month m + 1. The reason why this relation isn't satisfied automatically is that the adjustments and notional amounts are estimated at different times.

For example, the index may have generated a $5M redemption effective Jan. 06, but what was actually received from the funds is $4.9M. In the month prior to receiving the $4.9M the index would have calculated the outstanding residual amounts as $5M, and made adjustments based upon this value, even though it will turn out not to be accurate. If the index application were to try to correct for this inaccuracy, it would need to go back in time and regenerate adjustments retroactively, which it can't actually do in practice.

Even this sort of estimate inaccuracy shouldn't affect the sum rule, which is essentially conservation of value, except that the notional amounts are also typically estimates. When finalizing the index level, the cash inflow (the left side) in known since it represents client in-/out-flow that month. The right side represents the actual change in investments that actually took place in the index, so if we attempted to redeem $5M and only got $4.9M, there will definitely be a mismatch. This normalization process ensures that actual notional in-/out-flow experience is preserved, even if the net fund-level cash flows don't match it.

The normalization constant γm is therefore an interesting quantity to track to measure how accurate the index calculation estimates are.

The 'projected fund weights' are essentially an estimate of active fund weights that are used to determine the adjustments. This is essentially an identical formula to (17) with the new residual weight set to zero:

$$pfw^{bi}_{m+1,n} = (efw^{ei}_{m,n} - \varpi^{ei}_{m,n}) \frac{\eta^{e}_{m,n}}{\eta^{b}_{m+1,n}} + \sum_{k=1}^{m} \frac{\alpha^{i}_{m+1,k} \eta^{b}_{m+1,k}}{\eta^{b}_{m+1,k}}. \qquad (17)$$

Here *pfw* is the projected fund weight, *efw* the estimated fund weight, ̤ the residual weight, all quantities with a n indicating that these are calculated as of month n. These, then, are forward-looking estimates, but otherwise the calculation is identical to that described earlier.

Once the weights are calculated for a given month m in the future, we then need to determine what adjustments to make. the first step is to calculate the Aggregate Reallocation weight as:

$$arw_m = 1 - \sum_i pfw_m^i - \sum_i \varpi_m^{ei}. \tag{18}$$

The meaning of this is also clear. The estimates (17) take into account the expected change in index notional, but since there will have been no adjustments explicitly made for the current month, any notional inflow or outflow will not have been accounted for. If there in expected inflow in month m, then the sum of the projected fund weights will be less than one–not due to estimation errors in this case, but due to the notional increase, with no scheduled fund investments (yet).

So, on the right side of (18) is $\sum_i (pfw^i + \varpi^i)$, which is the estimate of the active weight from (18) plus the estimated residual amount (passive weight) for the month. This sum represents an estimate of the total amount 'invested' in the index, so one less the sum is the shortfall that must be allocated–or excess in the case of outflow. The rules (briefly discussed below) determine how the *arw* is spread out into allocations/redemptions to each fund.

Note that the projected fund weights depend on the reallocation weights/calculation on the previous month, so the calculations have to be performed as a bootstrap: projection, allocations (described below), then move to the next month.

The index references a proprietary database of hedge funds. The funds appearing in this database make up 'The Universe' of hedge funds as far as the index is concerned.

Since the index aims at being broadly representative of the hedge fund industry, it is important that The Universe represents a large enough fraction of the actual hedge fund universe.

Historically, this has been fairly difficult to determine, since not all funds will report NAVs and other information to one of the database vendors. Early estimates of statistics of the hedge fund world tended to be pretty inconsistent.

Every fund in The Universe is assigned to one of nine strategies (if possible), and a record of the total Assets Under Management ("AUM") in each strategy is maintained. The 12-month moving average of the AUM for each strategy is the Target Strategy Weight for the Index, capped by Target Strategy Weight Limits to avoid the index becoming, for example, effectively an equity long/short index. All this can be represented symbolically:

$$\text{TSW}_I = \min(\text{TSWL}_I, \text{AW}_I 12m\text{MA})) \qquad (19)$$

where I indicates the strategy, TSW the Target Strategy Weight, TSWL the Target Strategy Weight Limit, and AW(12mMA) the asset-weighted, 12 months moving average strategy weight in the Universe.

The Index will aim to have Strategy Weights ("SW") that track the strategies in the Universe by AUM, that is:

$$SW_I \sim TSW_I \qquad (20)$$

however, in practice SW and TSW will vary from month to month. When adjustments are made, they are made in such a way that (20) is approached (see [https://finpricing.com/lib/EqWarrant.html](https://finpricing.com/lib/EqWarrant.html)), but no specific adjustments related to re-balancing strategy weights are made unless:

$$SW_I \geq 1.2 \times TSW_I = Target\ Strategy\ Weight\ Cap_I \qquad (21)$$

3. Hedge Fund Index Adjustment

From the Universe, a subset of Eligible funds is determined. These are funds from the Universe that pass a number of Eligibility criteria: due diligence review, appropriate redemption and allocation terms, large enough AUM, etc. It may be that a fund is eligible and included in the index, only to become ineligible at a later time, which would trigger its removal.

The number of funds per strategy is initially set as:

$$N_I \sim 250 \times TSW_I \qquad (22)$$

If the number of funds in a strategy is less than this, then the rules attempt to add funds to the strategy so that this is satisfied. As the Index notional increases and funds in the index are capped, the index rules will add new funds to a strategy in order to attempt to satisfy (20).

Initially each fund in a given strategy will be equal-weighted, with weight: $TSW_I/N_I$. However, the weight of individual funds in the Index are limited by a few factors:

First, a global fund weight cap of 1%, which guarantees that in the worst-case scenario the index will be dominated by 100 funds. We feel that this cap is a little large, but this would only become a problem if there were a significant number of funds 'at' this cap.

Second, every fund has a $-exposure cap arising from exposure limits to every fund (there may also be a cap from the fund itself if it decides that it will not be accepting new allocations at the current time). Thea exposure limits apply to *all* exposures that the bank

has to the given fund, so the exposure cap for the Index will be reduced by non-Index-related exposure.

Third, a fund weight floor, which prevents the index from reducing its exposure to the fund. This can either arise from the fund's minimum investment requirements, or as a result of the funds closure (to new investments).

Note that fund caps take precedence over floors, so that if a cap is less than a floor a redemption is nevertheless triggered. the intent of the floor for closed funds is to avoid reducing exposure that cannot be regained–this is clearly an Index management strategy decision.

A noteworthy consequence of this is that a fund redemption that would otherwise be triggered by the rules (for example, as a consequence of the breach of a strategy or fund weight cap) may not be made immediately–the weights are projected forwards, and the best possible adjustments are made for every month subject to the redemption schedule of all the funds in the index. For example, if the rules say that we need to redeem from a Conv. Arb. fund due to a strategy weight cap breach, but no Conv. Arb. Fund in the index has a monthly liquidity (or we've missed the notice period), then it may be several months before a fund can be removed and the breach corrected.

This is performed 'Dutch auction' style, redeeming from the fund with the highest weight first, down to the level of the fund with the next highest weight. then these two funds would have their weights reduced to the next highest, and so on, until the necessary strategy weight is removed.

A recently discovered situation is where only a single fund was available for redemption, and the rules would have triggered a redemption from this single fund to an unreasonably small weight (or complete redemption), even though in the following months other funds would have been available for a more uniform redemption. The solution to this problem was to put a floor of 25bp for the weight of a fund that has had a redemption triggered through a strategy weight adjustment.

The reallocation weight is distributed amongst the strategies so as to minimize the difference between the weight strategies with shortfalls and the universe weights for those strategies ("Dutch auction" again).

If there is a shortfall in the target number of funds in a strategy (22), then funds are found to make up the difference–largest AUM first. Within a strategy, allocations are distributed to funds in a way that the smallest weights are as large as possible–subject to fund caps and floors, and with the restriction that this adjustment does not raise the weight above 75bp. This last restriction ensures that if there are only a few funds in the strategy that are available for allocation, they do not end up with unreasonably large weights.

If there is excess allocation to a strategy after the above is carried out, then new funds may be added to the strategy, or if this is not an option then the remainder is allocated to the other strategies.

The redemption weight is allocated to the strategies so that the maximum excess weight–amongst the strategies that have a weight in excess of its target weight–is as small as possible.

Within strategy redemptions are made from funds in the same manner as allocations, except that one is minimizing the largest fund weights, and it is possible for funds to be removed from the index. Not though, that not all funds will be available for redemptions on any given month.

4. Conclusion

This article presents a new model for constructing hedge fund index. The model is appropriate and found to be correctly implemented in the Index Application. This, however, is the easy part. The more difficult part is understanding how the index adjustment procedures work in detail, and it is here that we feel that there will be some ongoing fine-tuning as the index matures.

The determination of adjustments that need to be made at a particular point in time is not just for the current month but is projected forwards in time so that redemptions that need to be made are identified as early as possible, recalling that to redeem from a fund notice must be given. This makes the calculation of adjustments, e.g., fund allocation and redemption decisions, a complicated process. Since the calculations are based on estimates which are not recorded, it also means that it is quite difficult to retroactively verify that the adjustments are being made correctly.

The index has no mechanism to enable it to reduce exposure to a strategy other than the target strategy weight cap, such as setting at 120% of the target strategy weight. However, since the target strategy weights are 12 month moving averages of the strategy AUMs, they will respond to changes.

This means that if the hedge fund industry undergoes a fairly sudden shift–so that the AUM distribution amongst the strategies changes rapidly–the index may lag the universe. The index rules contain a provision to 're-start' the index, but there may be some tracking error relative to the universe in the meantime.

References:

Ardia, D. and K. Boudt (2018), "The Peer Performance Ratios of Hedge Funds", Journal of Banking & Finance, 87, 351-368.

Atilgan, Y., Bali, T., and Demirtas, K., 2013, The performance of hedge fund indices, Borsa Istanbul Review, 13 (3), 30-52.

Aragon, G. and V. Nanda (2017). Strategic delays and clustering in hedge fund reported returns. Journal of Financial and Quantitative Analysis 52, 1-35.

Barth, D., J. Joenväärä, M. Kauppila, and R. Wermers (2021). "The hedge fund industry is bigger (and has performed better) than you think." Working paper, Office of Financial Research.

Beschwitz, B., S. Lunghi, and D. Schmidt, 2022, Fundamental arbitrage under the microscope: evidence from detailed hedge fund transaction data, The Review of Asset Pricing Studies, 12 (1), 199-242.

Choi, James J and Kevin Zhao (2020). "Did Mutual Fund Return Persistence Persist? Working Paper 26707." National Bureau of Economic Research.

Dimmock, S., and W. Gerken (2016). "Regulatory oversight and return misreporting by hedge funds." Review of Finance 20, 795-821.

Hwang, I., S. Xu, F. In, and T. Kim (2017), Systemic Risk and Cross-Sectional Hedge Fund Returns, Journal of Empirical Finance, 42, 109-130.

Joenväärä, J., M. Kauppila, R. Kosowski, and P. Tolonen (2021), "Hedge fund performance: are stylized facts sensitive to which database one uses?" Critical Finance Review, Vol: 10, Pages: 271-327

Kapil, S. and J. Gupta (2019), Performance Characteristics of Hedge Fund Indices, Theoretical Economics Letters, 9 (6), 2176-2197.

Stafylas, D. and K. Anderson, 2018, Hedge fund index-engineering methodologies: a comparison and demonstration, Applied Economics, 50 (6), 596-612.